\documentclass[times, twoside]{Style_ArXiv}
\usepackage{blindtext}
\usepackage[overload]{textcase}

\begin{document}
    \bibliographystyle{Style_ArXiv}
    
\leadauthor{Bianconi}


\title{Giant Conductivity Modulation of Aluminum Oxide using Focused Ion Beam}
\shorttitle{Giant conductivity modulation}

\author[1]{Simone Bianconi}
\author[1,2]{Min-Su Park}
\author[1\space\Letter]{Hooman Mohseni}

\affil[1]{Department of Electrical and Computer Engineering, Northwestern University, Evanston, IL, 60208, USA}
\affil[2]{Nano Convergence Research Center
Korea Electronics Technology Institute
111 Ballyong-ro, Deokjin-gu, Jeonju 54853, South Korea}

\maketitle

\begin{abstract}

Precise control of the conductivity of semiconductors through doping has enabled the creation of advanced electronic devices, similarly, the ability to control the conductivity in oxides can enable novel advanced electronic and optoelectronic functionalities. While this was successfully shown for moderately insulating oxides, such as In\textsubscript{2}O\textsubscript{3}, a reliable method for increasing the conductivity of highly insulating, wide bandgap dielectrics, such as aluminum oxide (Al\textsubscript{2}O\textsubscript{3}), has not been reported yet.  Al\textsubscript{2}O\textsubscript{3} is a material of significant technological interest, permeating diverse fields of application, thanks to its exceptional mechanical strength and dielectric properties.
  
  Here we present a versatile method for precisely changing the conductivity of Al\textsubscript{2}O\textsubscript{3}. Our approach greatly exceeds the magnitude of the best previously reported change of conductivity in an oxide (In\textsubscript{2}O\textsubscript{3}). With an increase in conductivity of about 14 orders of magnitude, our method presents about 10 orders of magnitude higher change in conductivity than the best previously reported result. Our method can use focused ion beam to produce conductive zones with nanoscale resolution within the insulating Al\textsubscript{2}O\textsubscript{3} matrix. We investigated the source of conductivity modulation and identified trap-assisted conduction in the ion damage-induced defects as the main charge transport mechanism. Temperature-dependency of the conductivity and optical characterization of the patterned areas offer further insight into the nature of the conduction mechanism. We also show that the process is extremely reproducible and robust against moderate annealing temperatures and chemical environment.
  
The record conductivity modulation, combined with the nanoscale patterning precision allows the creation of conductive zones within a highly insulating, mechanically hard, chemically inert, and bio-compatible matrix, which could find broad applications in electronics, optoelectronics, and medical implants.

\end {abstract}

\begin{keywords}
    Thin Films | Alumina | Dielectrics | Defect Engineering | Conducting Oxides
\end{keywords}

\begin{corrauthor}
    hmohseni\at northwestern.edu
\end{corrauthor}

\section*{}
Aluminum oxide (Al\textsubscript{2}O\textsubscript{3}) is one of the most widely employed dielectric materials, thanks to its excellent insulating properties\cite{Belkin2017}, mechanical hardness and resistance\cite{Thamara2004}, and biocompatibility\cite{Thamara2004,Xifre2015}, with applications ranging from device passivation\cite{Belkin2017,Gaboriau2017,Dinge2010,Miya2010,Park2017,Xin2017}, MOSFET gate\cite{Hirama2007,Gutierrez2018,Zhang2016,Si2018,Peng2013}, to biomedical implants and antifouling passivation\cite{Thamara2004,Xifre2015,Dong2015,Feng2014}. Electrical functionalization of Al\textsubscript{2}O\textsubscript{3} via reliable and spatially-accurate control of its conductivity could enable novel sensing technologies encompassing electrical contacts embedded in a mechanically hard, chemically inert, and electrically insulating dielectric matrix. 
Examples of such technological applications include photon emission and detection\cite{Park2017,Mikhe2014,Wang2017,Jeske2017,Rezaei2017}, low-energy interconnects\cite{Ali2014,Miller1997,Miller2009}, energy conversion\cite{Hu2014,Liu2016,Belkin2017}, and implantable devices\cite{Kang2014,Fang2016,Diaz2017}.

We here present an effective method for the non-subtractive nanopatterning of electrically conductive wires and other features embedded in a dielectric Al\textsubscript{2}O\textsubscript{3} substrate, using focused ion beam (FIB).
While patterning of nanowires in transparent conductive In\textsubscript{2}O\textsubscript{3} has been demonstrated using FIB, the conductivity modulation was limited to 4 orders of magnitude\cite{Sosa2009,Sosa2010}. Here, we show nanopatterning of conductive zones in highly insulating Al\textsubscript{2}O\textsubscript{3}, achieving a conductivity modulation of 14 orders of magnitude (see Supporting Information)\cite{Dobro2009}. To the best of our knowledge, this is the highest reported change in conductivity using FIB.

FIB has proven to be an extremely effective tool for the nanopatterning of transparent conducting oxides (TCO) via lithographically controlled dopant implantation\cite{Sosa2009,Sosa2010}, as well as for structural modification and non-subtractive patterning of dielectric substrates\cite{Wang2017,Myers2012}. Despite being an inherently serial processing tool, FIB has shown very attractive potential in large-area milling and implantation, in particular when integrated with pattern generator lithography capabilities\cite{Palacios2010,Ocola2013,Ocola2014}. Furthermore, thanks to the higher beam deflection speed and lower settling time compared to electron beams, milling-based ion beam lithography (IBL) has demonstrated exposure times for large patterns that are comparable to those of EBL, and even faster patterning times can be achieved for ion implantation at lower dosage\cite{Imre2010}.

\section*{Methods}

The process schematic of our method is presented in Figure 1: the fabrication process is fairly simple, highly tunable, extremely versatile, and can be applied to both bulk Al\textsubscript{2}O\textsubscript{3} (i.e. sapphire) and thin film Al\textsubscript{2}O\textsubscript{3} (e.g. ALD). In order to limit the effects of charging, inherent to FIB processing of insulating substrates, a metallic alignment pattern is deposited on the sample before implantation using lithographic techniques (Figure 1a). This metallic film serves a dual purpose: it is connected to ground during implantation, allowing to avoid excessive charging of the substrate, and facilitates the focusing and alignment of the ion beam, crucial for the subsequent processing steps. Alternatively, in order to further improve the spatial resolution of the patterned nanowires, an anticharging thin layer of Au can be deposited over the whole area of the sample before implantation, and subsequently removed, as proposed by Sosa et al\cite{Sosa2009,Sosa2010}.

\begin{figure*}[t!]
  \centering
  \includegraphics[width=\textwidth]{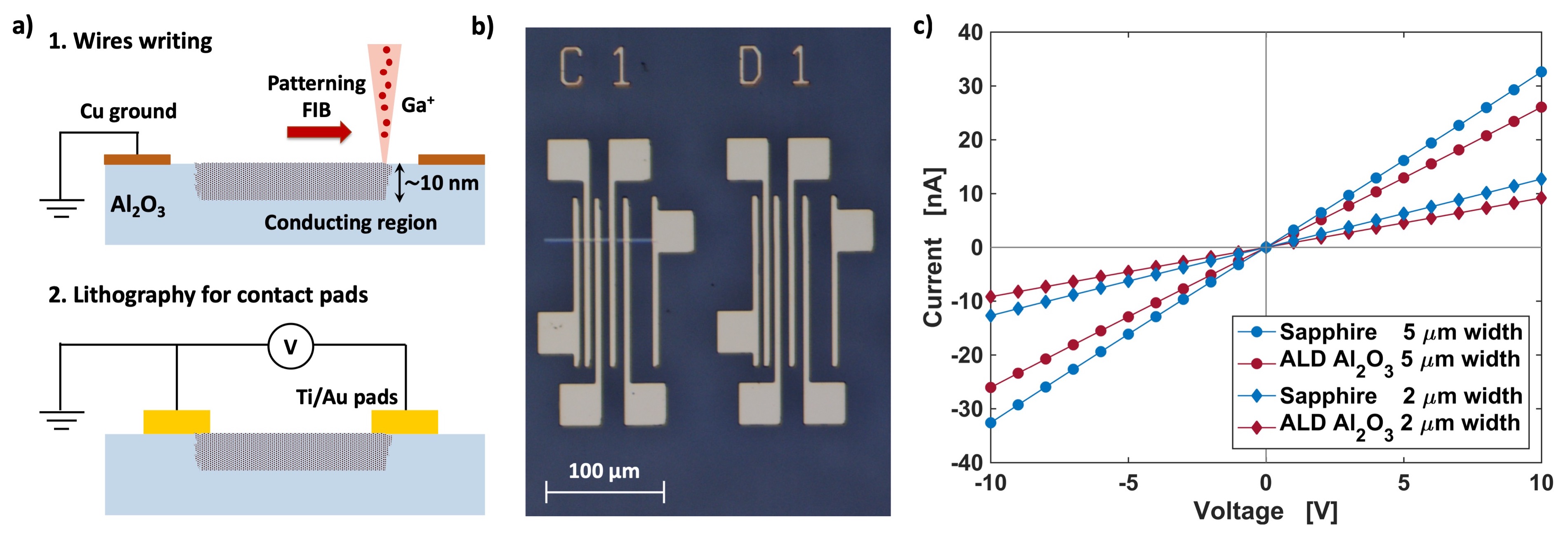}
  \caption{\textbf{Method for patterning of conductive wires in Al\textsubscript{2}O\textsubscript{3}. a)} Schematic of the patterning and fabrication of the wires in Al\textsubscript{2}O\textsubscript{3} matrix. \textbf{b)} Microscope image of the fabricated devices including the contact pads, and both implanted (left) and non-implanted devices (right). \textbf{c)} Ohmic current-voltage characteristics of patterned wires of varying widths implanted in bulk sapphire and ALD Al\textsubscript{2}O\textsubscript{3}.}
  \label{fig:1}
\end{figure*}

All the implanted wire-like structures exhibit ohmic behavior (Figure 1c). We argue that the mechanism behind this linear resistance is similar to the previously reported results from Ga-implanted TCO layers\cite{Sosa2010}. Ohmic conduction in dielectrics has been observed at low field for a variety of oxides and is related to weak carrier injection from the contacts\cite{Chiu2014,Wu2014}. We tested our method for patterning wires both in bulk sapphire substrates and ALD-deposited Al\textsubscript{2}O\textsubscript{3} films. As shown in Figure 1c, both matrices yield similar results in terms of the electrical conductivity, hence the rest of this paper focuses on the results from bulk sapphire for brevity. 

The electrical conductivity of the patterned nanowires is mainly determined by two factors: the geometrical size of the wires, and the ion dose. As shown in Figure 2a, the electrical conductivity scales linearly with the wire length. Furthermore, the high fidelity of the measurements over the large number of fabricated devices proves the reproducibility and robustness of this method (see Table S2 in the Supporting Information). The effect of implantation dose on the electrical characteristics of the wires is reported in Figure 2b: the conductivity of the nanowires varies by a few orders of magnitude within the range of doses explored. Notably, the conductivity reaches its maximum at a certain dose, and then plateaus and falls off at higher doses. Moreover, such peaks occur at higher doses for thinner wires, possibly due to the raster scanning of the patterning ion beam, that allows for a better mitigation of the charging effect in larger-area patterns. Nanopatterning of smaller wires is characterized by a more localized charging, which could deflect the incident ion beam and result in a lower effective dose delivered to the substrate\cite{Sosa2010}. 

\begin{figure*}[t!]
  \centering
  \includegraphics[width=0.8\textwidth]{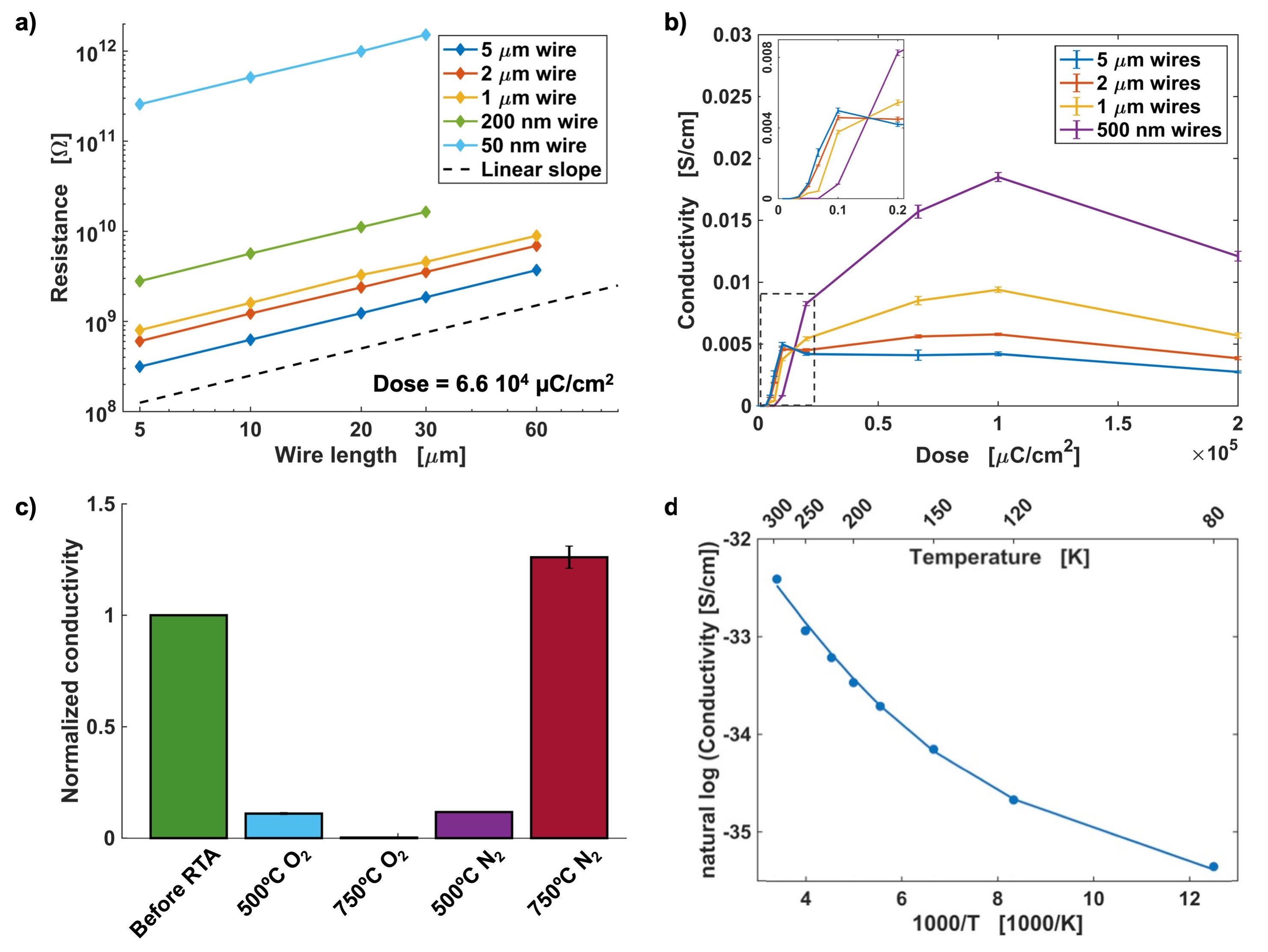}
  \caption{\textbf{Electrical characterization of the patterned wires. a)} Resistance scales with the wire length and width. \textbf{b)} Effect of ion dose on the wire conductivity. Error bars are calculated from a minimum of 5 devices per data point. \textbf{c)} Effect of RTA treatment on the conductivity of patterned Al\textsubscript{2}O\textsubscript{3} wires. Four different samples were treated with 15 s RTA in different environment and temperatures: the error bars are calculated from a minimum of 3 devices per data point. \textbf{d)} Arrhenius plot of the conductivity of patterned wires. The solid lines represent the fit used to estimate the activation energies, according to the model in eq.~\ref{eq:1}.}
  \label{fig:2}
\end{figure*}

\section*{Device Characterization}

\subsection*{Defect-assisted conduction}

Several theoretical and experimental studies have advanced the hypothesis that damage-induced defects in electronic materials can create deep trap states in the bandgap, which intrinsically act as dopants, enabling electrical conduction in otherwise highly insulating materials\cite{Sosa2010,Sosa2011,Choi2013,ChoiJanotti2013,Iberi2016,Madauss2016,Nanda2015}. These studies suggest that the electrical conductivity we observe in Al\textsubscript{2}O\textsubscript{3} could be due to the formation of oxygen vacancies caused by the ion collision cascade. To evaluate this hypothesis, we performed an annealing study as it was proved to be an effective tool for investigating the nature of the electrical conductivity in implanted wires in TCO\cite{Sosa2010}.

The results of rapid thermal annealing (RTA) treatment is presented in Figure 2c.  Annealing at 750$^\circ$C in an oxidizing environment effectively erases the implanted wire, reducing the electrical conductivity to negligible levels, while annealing in a reducing environment (N\textsubscript{2}) induces a higher conductivity in both the implanted wires and the non-implanted Al\textsubscript{2}O\textsubscript{3} matrix. RTA treatments performed at 500$^\circ$C result in a slight decrease in conductivity in both chemical environments, presumably due to incomplete activation of the defect annealing at such temperature. These results are consistent with published results for RTA treatment of implanted Ga nanowires in In\textsubscript{2}O\textsubscript{3}\cite{Sosa2010}, and suggest that the electrical conductivity of the patterned wires is related to ion damage-induced chemical and structural modification of the Al\textsubscript{2}O\textsubscript{3} matrix, rather than to the implanted Ga ions.

\subsection*{Charge transport}

To further evaluate the nature of the charge transport mechanism, we performed temperature-dependent conductivity characterization of the implanted devices. Figure 2d reveals a decrease in conductivity with the inverse of temperature. Transport in a wide bandgap dielectric such as Al\textsubscript{2}O\textsubscript{3} is governed by hopping conduction mechanisms of two types: direct trap-to-trap tunneling, and phonon-assisted elastic and inelastic tunneling\cite{Jegert2012,Callewaert2014,Kumar2014}. For both conduction mechanisms, temperature affects the density of free carriers, due to the thermal excitation of electrons at the trap sites\cite{Chiu2014}, and the carrier mobility, due to the thermally activated phonon scattering\cite{Wu2014}. Therefore, the temperature dependency of the conductivity $\sigma=nq\mu$ can be modeled as\cite{Chiu2014,Wu2014}:
\begin{equation}
    \sigma=\alpha  T^{-3/2}  e^\frac{-E_{ACT}}{k_B T}
    \label{eq:1}
\end{equation}
where $\alpha$ is a fitting parameter, $k_B$ is Boltzmann’s constant and $E_{ACT}$ is the activation energy in units of electron-Volt. Within the range of temperatures investigated, two conduction regimes are observed: one dominated by thermally activated phonons, at higher temperatures ($T > 200 K$), and one by direct hopping at lower temperatures; hence a good fit to the experimental data ($R^2=0.994$) is only possible by adding two functions of the form of eq.~\ref{eq:1}, with estimated activation energies of 96 meV and 50 meV.

\subsection*{Optical characterization}
We also evaluated the effect of the implantation on the optical properties of the sapphire substrate. Figure 3 shows a significant change in the photoluminescence (PL) and transmission spectra of implanted samples compared to the pristine sapphire. These measurements supporting the argument for the ion damage-induced formation of trap states within the oxide bandgap\cite{Townsend1994,Wu2002}. While the change in both PL and transmission due to implantation spans over a large range of wavelengths, interestingly, the characteristic $Cr^{3+}$ impurity peak shown in Figure 3b is unaffected by the implantation process, hence suggesting a wider energy distribution of trap states in the bandgap, as compared to atomic transition levels\cite{Kosty2016,Huang2014}.
\begin{figure*}[t!]
  \centering
  \includegraphics[width=0.8\textwidth]{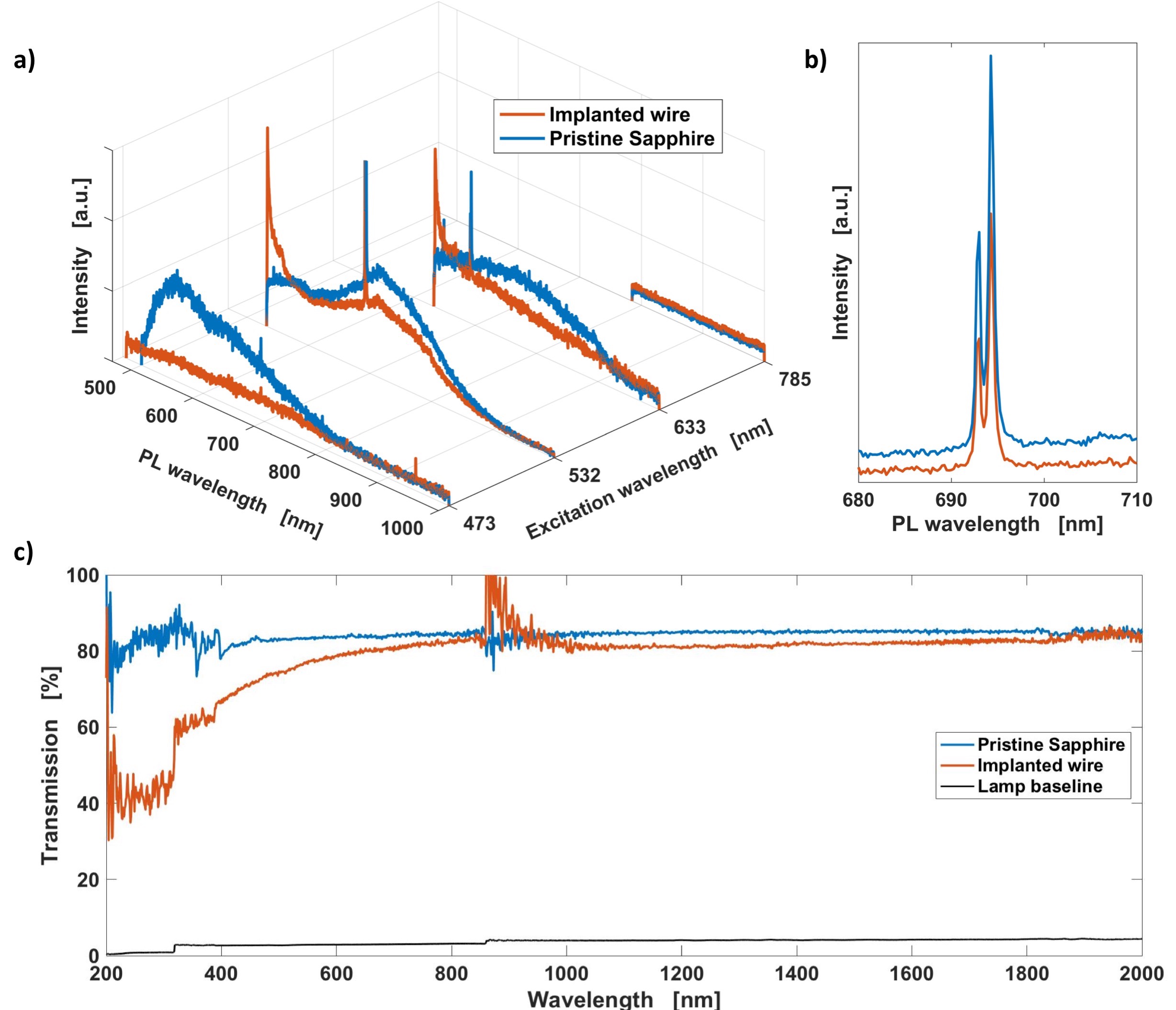}
  \caption{\textbf{Optical and chemical characterization of the patterned wires.} \textbf{a)} PL maps of implanted wire and pristine sapphire for different excitation wavelengths; \textbf{b)} Particular of the photoluminescence spectrum for 532 nm excitation wavelength, showing the $Cr^{3+}$ peaks for both implanted wire and pristine sapphire; \textbf{c)} Optical transmission as a function of wavelength for implanted wire and pristine sapphire.}
  \label{fig:3}
\end{figure*}

\section*{Conclusion}
We reported a novel method for nanopatterning of conductive zones in Al\textsubscript{2}O\textsubscript{3} matrix using focused ion beam irradiation. The implanted wires exhibit ohmic conduction with an average conductivity of $10^{-2}$ $S cm^{-1}$. Based on a large number of measured devices, we showed that the presented nanopatterning process has excellent uniformity and nanoscale spatial resolution, while allowing for modulation of the electrical conductivity through control of the ion dose. The electrical, optical, and chemical characterizations provide strong evidence that the conduction mechanism is due to the formation of trap states within the oxide bandgap. All processed devices showed stable performance during several months, and after heating to 100$^\circ$C and immersion in liquid solvents. The present method shows great potential for application in sensing technologies that require electrical contacts embedded in a mechanically hard, chemically inert, and electrically insulating dielectric matrix. Examples of such applications include photon sensors, low-energy interconnects, and micro/nano-implantable medical devices.

\section*{Experimental}
The majority of the devices presented were implanted in a 300 $\mu$m thick, double-side polished $<0001>$ crystalline sapphire substrates. Some of the devices were implanted in an Al\textsubscript{2}O\textsubscript{3} film deposited on both silicon and sapphire substrate using a Cambridge NanoTech Savannah S100 atomic layer deposition from Trimethylaluminum and water vapor precursors at 150$^\circ$C. An 80 nm Cu layer with alignment mark pattern was deposited on the sample before implantation using photolithographic lift-off techniques. The metallic layer was connected to ground during implantation. A FEI Nova 600 NanoLab dual-beam microscope with a Sidewinder Ga+ ion column and a 100 nm resolution piezomotor X-Y stage was employed for the wire implantation at ion beam energies of 5 and 30 keV and currents of 100 pA, 3.2 and 6.3 nA, yielding estimated spot sizes ranging 20 to 60 nm. The instrument was combined with a RAITH Elphy 4.0 interface with integrated 16-bit DAC pattern generator, which enables large area patterning and write-field alignment, accounting for stitching and drift, and allows for precise dose control via beam blank and deflection. The Elphy interface allows for a multi-pass patterning technique, which help contrast the strong charging in dielectrics. Nevertheless, we estimate the effective dose to be lower than the nominal value, due to sputtering and localized charging deflecting the incident ions. The Ti/Au contacts were subsequently deposited with photolithographic lift-off techniques, and electrical testing was performed with a 4-point probe station equipped with Agilent 4285A LCR meter. Photoluminescence spectra were acquired using a HORIBA LabRAM HR Evolution confocal Raman, equipped with 473 nm, 532 nm, 633 nm and 785 nm lasers for excitation.

\begin{acknowledgements}
 The authors acknowledge partial support from ARO award $\#$W911NF-18-1-0429. This work was performed, in part, at the Center for Nanoscale Materials of Argonne National Laboratory. Use of the Center for Nanoscale Materials, an Office of Science user facility, was supported by the U.S. Department of Energy, Office of Science, Office of Basic Energy Sciences, under Contract No. DE-AC02-06CH11357. S.B. gratefully acknowledges support from the Ryan Fellowship and the International Institute for Nanotechnology at Northwestern University.
\end{acknowledgements}

\begin{contributions}
  The manuscript was written through contributions of all authors. All authors have given approval to the final version of the manuscript.
\end{contributions}

\begin{interests}
 The authors declare no competing financial interests.
\end{interests}

\begin{suppinfo}
The following supporting information is available from the author:

\textbf{Table S1:} Calculations of magnitude conductivity modulation in Al\textsubscript{2}O\textsubscript{3} reported in this paper compared to that previously reported for In\textsubscript{2}O\textsubscript{3}\cite{Sosa2011,Sosa2010,Sosa2009}.\\
\textbf{Figure S1:} Additional proof for uniformity of the implanted conductive regions\\
\textbf{Figure S2:} Additional graphic visualization of current-voltage characteristics after RTA treatments\\
\textbf{Table S2:} Complete set of current-voltage measurements performed on all the fabricated devices

\end{suppinfo}

\section*{Bibliography}
\bibliographystyle{unsrt}
\bibliography{Giant_biblio}

\end{document}